# Trees and Forests for Nonequilibrium Purposes: An Introduction to Graphical Representations

Faezeh Khodabandehlou[1] · Christian Maes[1] · Karel Netočný[2]

**Abstract**
Using local detailed balance we rewrite the Kirchhoff formula for stationary distributions of Markov jump processes in terms of a physically interpretable tree-ensemble. We use that arborification of path-space integration to derive a McLennan-tree characterization close to equilibrium, as well as to obtain response formula for the stationary distribution in the asymptotic regime of large driving. Graphical expressions of currents and of traffic are obtained, allowing the study of various asymptotic regimes. Finally, we present how the matrix-forest theorem gives a representation of quasi-potentials, as used e.g. for computing excess work and heat in nonequilibrium thermal physics. A variety of examples illustrate and explain the graph elements and constructions.

## 1 Introduction

Equilibrium statistical mechanics describes the fluctuations in a system of many degrees of freedom upon imposing certain macroscopic conditions. A typical scenario is to give the energy $E(x)$ as function of the state $x$ of the system, e.g., a sum of local energies built from internal potentials and interactions, and the temperature $\beta^{-1}$ of the surrounding heat bath. From the Boltzmann-Gibbs weights $\propto \exp -\beta E(x)$ for state $x$ also the static macroscopic fluctuations in equilibrium can be obtained as governed by thermodynamic potentials.

Unsurprisingly, there is no such unifying Gibbs algorithm for nonequilibrium (driven or active) systems. For one, the dynamics starts to matter much more, even for the static (stationary) fluctuations. For example, when the system is subject to important differences in temperature or chemical potential at its boundary, or when strong bulk driving is present, or under active forces, the stationary probabilities may very well depend on kinetic details such as the way in which energy gets exchanged with thermal baths or on diffusion properties, etc. One possible approach is to work instead with *dynamical* ensembles [1], i.e., with probability distributions on path-space given the statistics of possible system trajectories. In these dynamical ensembles, we integrate over all allowed paths, where the weights are expressed in terms of an action on paths that, in the case of local detailed balance [2–10], allows a decomposition in an entropic/dissipative part and a frenetic/nondissipative part; see [11, 12]. We know for example, since the paper by Komatsu and Nakagawa [13] how to connect that path-space integration with static fluctuations. That emphasis on path-space also fits with up-to-date experimental progress in manipulating and visualizing trajectories in mesoscopic systems, e.g. for colloids and micro-rheology or in biophysical processes. Yet, the full trajectory-picture, e.g. for processes described via Markov jump processes, mixes activity and dissipation, waiting and hopping, and the space of trajectories is obviously infinite-dimensional. An intermediate description that is theoretically a possible alternative, may be offered by graphical representations in terms of trees and forests. We have in mind the *arborification* of trajectories as suggested in e.g. [14–17]. To review and extend these techniques is the subject of the present paper. As a *bonus* we observe that these graphical tree-like representations may lead to important simplifications in certain asymptotic regimes (low temperature, large driving).

Markov jump processes on finite connected graphs allow their stationary distribution, solution of the stationary Master equation, to be expressed via the so called Kirchhoff formula. The origin of that formula is in the global balance between sink and source terms, specifying stationarity much in the same way as for linear electric circuits. In mathematical terms, the formula results from the matrix-tree theorem and has a natural extension in the matrix-forest theorem that we discuss as well. More generally, we explain and partially review how graphical representations shed light on nonequilibrium issues in the (restricted) context of Markov jump processes. It is worth mentioning here that the combination of nonequilibrium aspects with graph theory has possibly a wider application area than physics alone.

*Plan of the paper* We start with introducing a tree-ensemble, which gives a graphical representation of the stationary distribution, which is useful for deriving heat bounds and for easy access to various physical regimes of discussion. We explain in Sect. 3 its advantage in the far-from-equilibrium regime, such as for response theory. The Kirchhoff formula of Sect. 2 indeed simplifies and becomes physically interesting in certain asymptotic regimes. Section 4 is devoted to graphical expressions for currents and dynamical activity, and those are especially helpful in the low-temperature asymptotics. In Sect. 5 we describe the arborification of the McLennan ensemble, which summarizes the close-to-equilibrium regime in tree-language. Finally, in Sect. 6 we present an application of the matrix-forest theorem to construct so called quasi-potentials, [18], used to characterize excesses in heat and work during the relaxation from one to another nonequilibrium steady condition.

Throughout the paper, we use simple examples, which can be skipped at any time, to illustrate points which are much more general and easily visible from the graph representations.

## 2 Tree Ensemble

We consider a continuous-time Markov random walk on a finite connected graph $\mathcal{G}$. In such modeling, the vertices usually represent (physically coarse-grained) states of an open system, e.g, giving the position of particles such as colloids or ions in lattice gases or summarizing the chemo-mechanical configuration of molecular motors. Discreteness or the digital nature of dynamical activity, sometimes more appropriate for quantum descriptions, is reflected in the 'jump'- character of the transitions.

Transitions are only possible between vertices (states) $x, y$ which are linked by an edge in the graph, and then have a transition rate $k(x, y)$ for $x \to y$. We parametrize those rates as to allow an easy interpretation of local detailed balance [2–10],

$$k(x, y) = \psi(x, y) \, e^{S(x,y)/2} \qquad (2.1)$$

with symmetric reactivities $\psi(x, y) = \psi(y, x) > 0$, nonzero only when $\{x, y\}$ is an edge in $\mathcal{G}$, and with antisymmetric entropy fluxes $S(x, y) = -S(y, x)$: for a jump $x \to y$ over the directed edge $(x, y)$ in the graph, the $S(x, y)$ gives the corresponding change of entropy per $k_B$ in one of the equilibrium (electro-)chemical or thermal reservoirs making the environment. Breaking of detailed balance is mathematically visible from the nongradient or rotational character of the entropy fluxes, e.g., that there exist cycles $x_1 \to x_2 \to \ldots x_n \to x_1$ on the graph, $\psi(x_i, x_{i+1}) \neq 0$, for which

$$S(x_1, x_2) + S(x_2, x_3) + \ldots + S(x_{n-1}, x_n) + S(x_n, x_1) \neq 0 \qquad (2.2)$$

Note that when there is a potential $V$ so that for all transitions $S(x, y) = \beta(V(x) - V(y))$ (as for -global- detailed balance in a thermal medium at temperature $T$, $k_B T = \beta^{-1}$ where from now we set $k_B = 1$), then (2.2) equals zero.

## 2.1 Kirchhoff Formula

An expression for the stationary probability distribution, i.e., for the unique normalized solution $\rho^s > 0$ of the stationary Master equation,

$$\sum_y [k(x, y)\rho(x) - k(y, x)\rho(y)] = 0, \quad \text{for all } x$$

can be given from Kirchhoff's formula, [19–21]. First define a spanning tree as a tree that includes all vertices of the graph. Let $\mathcal{T}$ be such a spanning tree; we write $\mathcal{T}_x$ for the in-tree to $x$, obtained by orienting every edge in $\mathcal{T}$ towards $x$. Then,

$$\rho^s(x) = \frac{1}{W} w(x), \quad w(x) = \sum_\mathcal{T} \omega_\mathcal{T}(x), \quad \omega_\mathcal{T}(x) = W(\mathcal{T}_x) := \prod_{(z,z') \in \mathcal{T}_x} k(z, z') \qquad (2.3)$$

where $W = \sum_x w(x) = \sum_{\mathcal{T},x} W(\mathcal{T}_x)$ is the normalization. A directed edge from $x$ to $y$ is shown by $(x, y)$ and $\{x, y\}$ denotes an undirected edge between states $x$ and $y$. We often write $k_{xy} = k(x, y)$, $k_{zv} = k(z, v)$ etc as shorthand.

Just to make sure about the notation, we review two very easy examples.

**Example 2.1** Consider the graph in Fig. 1.

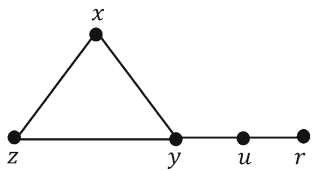

**Fig. 1** A loop over three states $x, y, z$ and attached hair $y - u - r$

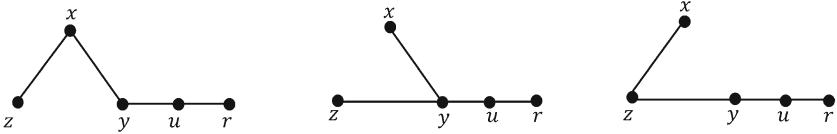

**Fig. 2** Spanning trees for the graph in Fig.1

The spanning trees are drawn in Fig. 2.
and

$$w(x) = \left(k_{yx}k_{zy} + k_{zx}k_{yx} + k_{yz}k_{zx}\right) k_{ru}k_{uy}$$
$$w(y) = \left(k_{zx}k_{xy} + k_{zy}k_{xy} + k_{xz}k_{zy}\right) k_{ru}k_{uy}$$
$$w(z) = \left(k_{yx}k_{xz} + k_{xy}k_{yz} + k_{xz}k_{yz}\right) k_{ru}k_{uy}$$
$$w(u) = \left(k_{zx}k_{xy} + k_{zy}k_{xy} + k_{xz}k_{zy}\right) k_{ru}k_{yu}$$
$$w(r) = \left(k_{zx}k_{xy} + k_{zy}k_{xy} + k_{xz}k_{zy}\right) k_{ur}k_{yu}$$

**Example 2.2** As a second example we consider Fig. 3.

The spanning trees are given in Fig. 4.

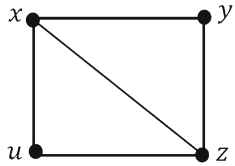

**Fig. 3** Graph with overlapping loops

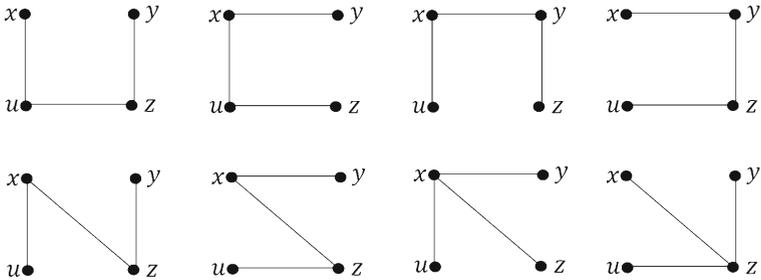

**Fig. 4** Spanning trees for Fig. 3

and

$$\begin{aligned}
w(x) &= k_{yz}k_{zu}k_{ux} + k_{uz}k_{zy}k_{yx} + k_{zu}k_{ux}k_{yx} + k_{zy}k_{yx}k_{ux} \\
&\quad + k_{yz}k_{zx}k_{ux} + k_{yx}k_{uz}k_{zx} + k_{ux}k_{zx}k_{yx} + k_{yz}k_{zx}k_{uz} \\
w(y) &= k_{xu}k_{uz}k_{zy} + k_{uz}k_{zy}k_{xy} + k_{zu}k_{ux}k_{xy} + k_{ux}k_{xy}k_{zy} \\
&\quad + k_{ux}k_{xz}k_{zy} + k_{uz}k_{zx}k_{xy} + k_{ux}k_{zx}k_{xy} + k_{xz}k_{uz}k_{zy} \\
w(z) &= k_{xu}k_{uz}k_{yz} + k_{uz}k_{xy}k_{yz} + k_{yx}k_{xu}k_{uz} + k_{ux}k_{xy}k_{yz} \\
&\quad + k_{ux}k_{xz}k_{yz} + k_{yx}k_{xz}k_{uz} + k_{yx}k_{ux}k_{xz} + k_{xz}k_{yz}k_{uz} \\
w(u) &= k_{yz}k_{zu}k_{xu} + k_{xy}k_{yz}k_{zu} + k_{yx}k_{xu}k_{zu} + k_{zy}k_{yx}k_{xu} \\
&\quad + k_{yz}k_{zx}k_{xu} + k_{yx}k_{xz}k_{zu} + k_{yz}k_{xz}k_{zu} + k_{yx}k_{zx}k_{xu}
\end{aligned}$$

## 2.2 Stationary Distribution from Dissipation on Trees

The representation (2.3) does not require any physical input; the physics is largely irrelevant in writing (2.1). That can be helped: we continue by giving an equivalent representation of the stationary distribution where the dissipation plays a central role.

Let $\mathcal{T}$ be a spanning tree of the graph $\mathcal{G}$. On that tree $\mathcal{T}$, there is a unique path from $y$ to $x$. All states $u$ on that path can have some rooted sub-trees ( "leaves") towards that $u$. The leaves are indicated by $\mathcal{L}$ in Fig. 5.

Local detailed balance (2.1) implies that for each spanning tree $\mathcal{T}$,

$$\frac{\omega_{\mathcal{T}}(x)}{\omega_{\mathcal{T}}(y)} = e^{\frac{1}{2}[S(\mathcal{T}_x) - S(\mathcal{T}_y)]} \tag{2.4}$$

where $S(\mathcal{T}_x) := \sum_{e \in \mathcal{T}_x} s(z, z')$ is the sum over the edges $e = (z, z')$ in the in-tree $\mathcal{T}_x$. Consider now the unique path between $y$ and $x$ in the spanning tree $\mathcal{T}$; see Fig. 5. Then,

$$S(\mathcal{T}_x) = S(\mathcal{T}_y) + 2S_{\mathcal{T}}(y \to x) \tag{2.5}$$

where $S_{\mathcal{T}}(y \to x)$ is the sum of the entropy fluxes $S(u, v) = \log \frac{k(u,v)}{k(v,u)}$ in $\mathcal{T}$ when moving on that tree from $y$ to $x$. By antisymmetry, $S_{\mathcal{T}}(y \to x) = -S_{\mathcal{T}}(x \to y)$ and hence (2.4) implies

$$\frac{\omega_{\mathcal{T}}(x)}{\omega_{\mathcal{T}}(y)} = e^{S_{\mathcal{T}}(y \to x)} \tag{2.6}$$

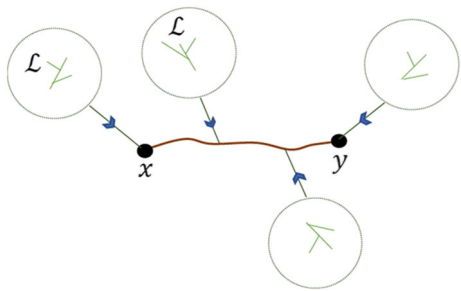

**Fig. 5** The tree $\mathcal{T}$ and the path between $x$ and $y$ with the leaves coming to vertices on that path

which is the tree-version of local detailed balance (2.1). As check, in the case of — global — detailed balance where $S(u, v) = U(u) - U(v)$ for some $U$, we always have $S_\mathcal{T}(y \to x) = U(y) - U(x)$.

Next, we define the tree-ensemble (depending on a state $x$)

$$\langle \cdot \rangle_x = \frac{1}{w(x)} \sum_\mathcal{T} \cdot \, \omega_\mathcal{T}(x) \tag{2.7}$$

which defines a probability measure on trees. From (2.3) and (2.6) follows

$$\frac{\rho^s(y)}{\rho^s(x)} = \frac{w(y)}{w(x)} = \langle e^{S_\mathcal{T}(x \to y)} \rangle_x \tag{2.8}$$

and summing over $y$ gives the more physical graphical representation of the stationary distribution,

$$\frac{1}{\rho^s(x)} = \sum_y \langle e^{-S_\mathcal{T}(y \to x)} \rangle_x \tag{2.9}$$

A similar expression for the stationary distribution can be found in [13], but here we express $\rho^s$ in terms of trees instead of via the path-space measure. Such a type of "arborification" has been mentioned as a possibility in [15]. Note that from (2.8) we get an improvement of the heat bounds derived in [14], as inequalities are readily obtained. Indeed we wish to emphasize that (2.9) is nonperturbative. Yet, it is a very natural starting point for e.g. understanding the linear regime close-to-equilibrium, as will be carried on in Sect. 5 for obtaining the tree–McLennan formula.

To recapitulate, formula (2.9) also implies the Kirchhoff formula (2.3) as can be seen by the chain of equalities,

$$\sum_y \langle e^{S_\mathcal{T}(x \to y)} \rangle_x = \sum_y \sum_\mathcal{T} \frac{\omega_\mathcal{T}(x)}{w(x)} e^{-S_\mathcal{T}(y \to x)} = \sum_y \frac{1}{w(x)} \sum_\mathcal{T} \omega_\mathcal{T}(x) e^{-S_\mathcal{T}(y \to x)}$$

$$= \sum_y \frac{1}{w(x)} \sum_\mathcal{T} \omega_\mathcal{T}(y) = \frac{W}{w(x)} = \frac{1}{\rho^s(x)}$$

We have used (2.6) here to reach the second line.

## 3 Far-from-Equilibrium Stationary Distribution

The present section transforms general results for Markov jump processes included in [14, 17] concerning low-temperature nonequilibria into an asymptotics far away from detailed balance. That yields a physical-kinetic interpretation of what nonequilibrium conditions get highest probability when the driving is very high. Again, the point of departure is the Kirchhoff formula, but turned into a physically interpretable expression.

We write $S(x, y) = \beta(V(x) - V(y)) + \mathcal{E} s(x, y)$ and $\psi(x, y) = \psi_\mathcal{E}(x, y)$ with $\mathcal{E} \gg 1$ in (2.1) to quantify the far-from-equilibrium condition. We think of $\mathcal{E}$ as distance from equilibrium, e.g. from the consumption of ATP in biochemical processes. making some loop-contributions (2.2) large. Note that we typically have $\mathcal{E} s(x, y) = \beta w(x, y)$ where $w(x, y) = -w(y, x)$ is the irreversible work done on the system during $x \to y$, and $\beta$ the inverse temperature of an ambient thermal bath. Large driving then means fixing $\beta$ and having $w(x, y) \uparrow \infty$.

To have that background driving amplitude $\mathcal{E}$ enter the transition rates (2.1) explicitly, we rewrite (2.1) into

$$k(x, y) = a(x, y) \, e^{-\mathcal{E}[\Gamma(x)+U(x,y)]} \quad (3.1)$$

where $a(x, y) = e^{O(\mathcal{E})}$ is sub-exponential in $\mathcal{E} \uparrow \infty$, and depends most often on the reactivity $\psi(x, y)$ and on the irrotational fluxes $\beta[V(x) - V(y)]$. In (3.1) enters minus the log-asymptotic escape rate $\Gamma$ for which $\sum_y k(x, y) \simeq e^{-\mathcal{E}\Gamma(x)}$, meaning

$$\Gamma(x) := -\lim_{\mathcal{E}} \frac{1}{\mathcal{E}} \log \sum_y k(x, y) = -\max_y \lim_{\mathcal{E}} \frac{1}{\mathcal{E}} \log k(x, y)$$

and the log-asymptotic transition probability $e^{-\mathcal{E} U(x,y)}$, meaning

$$U(x, y) = -\Gamma(x) - \lim_{\mathcal{E}} \frac{1}{\mathcal{E}} \log k(x, y) \geq 0 \quad (3.2)$$

which plays the role of a cost function over the edge $x \to y$.

In general, states $y$ with $U(x, y) = 0$ are called the *preferred successor*s of $x$ and always exist. From there follows the digraph $\mathcal{G}^{\text{dir}}$ having the same set of states as $\mathcal{G}$ but where the edges get oriented to point from one state to all its preferred successors. Therefore, directed graphs or digraphs are the architecture of strongly driven nonequilibrium Markov jump processes.

Remark that comparing (3.1) with (2.1), the nonequilibrium amplitude $\mathcal{E} \gg 1$ may also influence the time-symmetric reactivities $\psi(x, y) = \psi_\mathcal{E}(x, y)$, which is responsible for important physical effects such as jamming or negative differential conductivities [22]. However, if $\psi(x, y)$ is only subexponential in $\mathcal{E}$, then $\Gamma(x) = -\max_y s(x, y)/2$ and the escape rate is large at $x$ when a highly dissipative channel is available.

With the parametrization (3.1) and following [17], we proceed to give the asymptotic expression of the stationary distribution $\rho^s$ for $\mathcal{E} \gg 1$.

Let $\mathcal{T}$ be a spanning tree of $\mathcal{G}$ and $\mathcal{T}_x$ the in-tree to $x$ by orienting every edge in $\mathcal{T}$ towards $x$. Define

$$\Theta(x) := \min_{\mathcal{T}} U(\mathcal{T}_x); \quad U(\mathcal{T}_x) := \sum_{(y,y') \in \mathcal{T}_x} U(y, y')$$

as minimal cost to globally reach the condition $x$. We put $M(x)$ to be the set of spanning trees $\mathcal{T}$ for which that minimum is reached: $\mathcal{T} \in M(x)$ if $\Theta(x) = U(\mathcal{T}_x)$. Finally, let

$$A(x) := \sum_{\mathcal{T} \in M(x)} \prod_{(y,y') \in \mathcal{T}_x} a(y, y') \quad (3.3)$$

taking the product of the subexponential part in the symmetric reactivities; see (3.1). Then, by the same arguments as used in [17], there is a $\delta > 0$ so that

$$\rho^s(x) = \frac{1}{\mathcal{Z}} A(x) \, e^{\mathcal{E}[\Gamma(x) - \Theta(x)]} (1 + O(e^{-\delta \mathcal{E}})) \quad (3.4)$$

The notation $O(e^{-\delta \mathcal{E}})$ indicates a correction term which goes to zero exponentially fast with $\mathcal{E} \uparrow \infty$ at some positive rate $\delta$. In particular, the log-asymptotics of the stationary distribution (3.4) reads

$$\rho^s(x) \simeq e^{-\mathcal{E}(\Phi(x) - \min_y \Phi(y))}, \quad \Phi(x) := \Theta(x) - \Gamma(x)$$

The interpretation is more kinetic than thermodynamic: we already know that $\exp \mathcal{E}\Gamma$ is a lifetime, and we can interpret $\exp -\mathcal{E}\Theta$ as an accessibility. The pseudo-potential $\Phi$ governing the stationary distribution $\rho^s$ for large $\mathcal{E}$ thus combines the weights of life-time and accessibility: far-from-equilibrium, the conditions $x$ with lowest escape rate and that are most easily accessible get the largest occupation.

We call
$$\mathcal{D} = \{x^* : \Phi(x^*) = \min_x \Phi(x)\} \quad (3.5)$$
the set of dominant states, in the sense that they dominate in the far-from-equilibrium regime. Among them and following from (3.4), the most probable state is decided by the prefactor $A$ of (3.3), possibly also still depending on $\mathcal{E}$.

### 3.1 Examples

For illustration we collect a number of elementary examples.

**Example 3.1** Suppose a random walk on the ring $\mathbb{Z}_n$ with $n$ states and where $\mathcal{E}$ is the bias (external field) for nearest neighbor hopping with rates
$$k(x, x+1) = a_x\, e^{\mathcal{E}/2}, \quad k(x+1, x) = a_x\, e^{-\mathcal{E}/2} \quad (3.6)$$
so that in (3.2) we have $U(x, x+1) = 0, U(x+1, x) = 1, \Gamma(x) = -1/2$. All states are dominant.

Following (3.4) the stationary distribution is
$$\rho^s(x) \propto \frac{1}{a_x}, \quad x \in \mathbb{Z}_n$$
completely determined (always in the regime $\mathcal{E} \gg 1$) by the symmetric activities $\psi(x, y)$. Hence, those activities (and in particular the lowest escape rate $a_x\, e^{\mathcal{E}/2} + a_{x-1}\, e^{-\mathcal{E}/2}$ for large $\mathcal{E}$) decide the most probable state among the dominant states. It is an instance of the so called (Landauer) blowtorch theorem (1975), [23], that we can use those activities to change the relative occupations; see [14].

**Example 3.2** Consider again a random walk on the ring $\mathbb{Z}_n$ as in the previous example but now with the rates
$$k(x, x+1) = e^{\beta \mathcal{E}/2} e^{\beta(V(x)-V(x+1))/2}, \quad k(x, x-1) = e^{-\beta \mathcal{E}/2} e^{\beta(V(x)-V(x-1))/2} \quad (3.7)$$
where we have introduced the inverse temperature $\beta$ of an ambient thermal bath. We consider then the case of large $\beta$, i.e., the low-temperature asymptotics, which formally can be treated similarly to large driving (while we fix $\mathcal{E} = 1$).

Clearly, the state with the smallest value for the potential $V$ is not always the dominant state. However, we can always make a state dominant by lowering its value for the potential. In Figs. 6–7 we consider $\mathbb{Z}_7$ (seven states on a ring). We plot $V(x)$ and $e^{-\beta \Phi(x)}$ where $x = 1, 2, .., 7$ and $\beta = 1$ (see (3.4)).

The figures are illustrating the point that by choosing a sufficiently small value for the potential of a specific state we can make that state (uniquely) dominant. To show, let us fix a state $x$ and see how for given values $V(y), y \neq x$, we can always find a small enough value $V(x)$ so that $x$ becomes dominant, i.e., so that the effective potential in (3.4) verifies $\Phi(x) < \Phi(y)$ for all $y \neq x$.

The strategy is as follows. Define
$$\phi(z, z') := \lim_{\beta \to \infty} \frac{1}{\beta} \log k(z, z') \quad \text{and} \quad \phi(\mathcal{T}_x) := -\sum_{(z,z') \in \mathcal{T}_x} \phi(z, z')$$

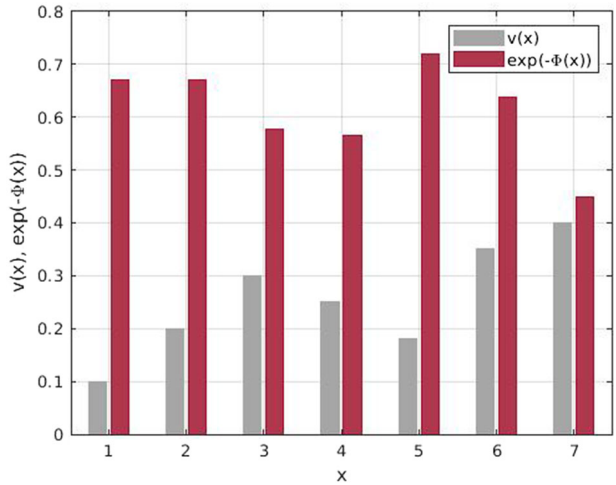

**Fig. 6** The smallest potential is $V(1)$ but the state $x = 5$ is the dominant state

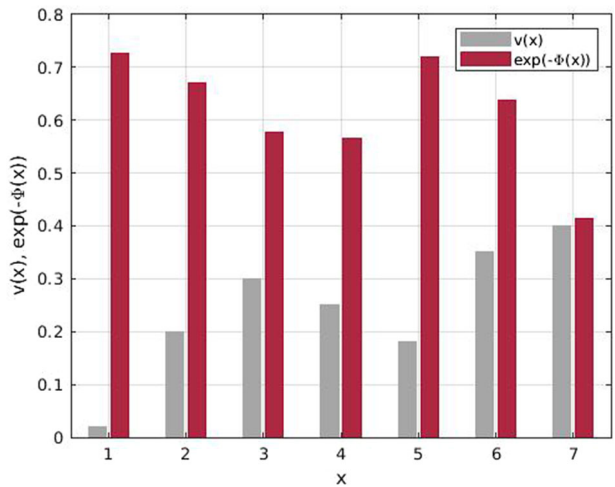

**Fig. 7** The smallest potential is $V(1)$ and the state $x = 1$ is the dominant state

in which $\mathcal{T}_x$ is a spanning in-tree to state $x$. Then,

$$\Phi(x) < \Phi(y) \quad \text{iff} \quad \forall \mathcal{T} \in M(x), \mathcal{T}' \in M(y), \quad \phi(\mathcal{T}_x) < \phi(\mathcal{T}'_y) \quad (3.8)$$

It suffices therefore that we can make $V(x)$ so small that for all $\mathcal{T} \in M(x), \mathcal{T}' \in M(y), \phi(\mathcal{T}_x) < \phi(\mathcal{T}'_y)$. To prove (3.8), we first show that

$$\forall \mathcal{T}^2, \mathcal{T}^1 \in M(x), \quad \phi(\mathcal{T}^1_x) = \phi(\mathcal{T}^2_x)$$

The reason is that

$$\phi(\mathcal{T}_x^1) = -\sum_{(z,z')\in\mathcal{T}_x^1}\Gamma(z) + U(\mathcal{T}_x^1)$$
$$= -\sum_{(z,z')\in\mathcal{T}_x^1}\Gamma(z) + U(\mathcal{T}_x^2)$$
$$= -\sum_{z\in\mathbb{Z}_n}\Gamma(z) + \Gamma(x) + U(\mathcal{T}_x^2)$$
$$= -\sum_{(z,z')\in\mathcal{T}_x^2}\Gamma(z) + U(\mathcal{T}_x^2) = \phi(\mathcal{T}_x^2)$$

But for all $\mathcal{T} \in M(x)$,

$$\Phi(x) = -\Gamma(x) - \sum_{z\neq x}\Gamma(z) - \sum_{(z,z')\in\mathcal{T}_x}\phi(z,z') = -\sum_{z\in\mathbb{Z}_n}\Gamma(z) + \phi(\mathcal{T}_x)$$

and for all $\mathcal{T}' \in M(y)$,

$$\Phi(y) = -\sum_{z\in\mathbb{Z}_n}\Gamma(z) + \phi(\mathcal{T}'_y)$$

In fact, the above reasoning holds true for every digraph.

To finally prove (3.8), we simply consider all spanning in-trees to $x$. They come in three different types:

$$\mathcal{T}_x^1 = (x+1, x+2, ..., x-1, x)$$
$$\mathcal{T}_x^2 = (x-1, x-2, ..., x+1, x)$$
$$\mathcal{T}_x^3 = (z, z-1, ..., x-1, x; z+1, z+2, ..., x-1, x); \quad z \in \mathbb{Z}_n, z \neq x$$

and

$$\phi(\mathcal{T}_x^1) = -(\phi(x+1) + \phi(x+2) + ... + \phi(x-1) + \phi(x))$$
$$= \frac{1}{2}(V(x) - V(x+1)) - \frac{n-1}{2}$$
$$\phi(\mathcal{T}_x^2) = \frac{1}{2}(V(x) - V(x-1)) + \frac{n-1}{2}$$
$$\phi(\mathcal{T}_x^3) = \frac{1}{2}(2V(x) - V(z) - V(z+1)) + \frac{k-m}{2}, \quad m+k = n-1$$

where $m$ and $k$ are respectively the number of edges from $z$ to $x$ and from $z+1$ to $x$. Obviously, the same scenario applies for state $y$, giving three different types of $\phi(\mathcal{T}_y)$. Making a comparison between $\phi(\mathcal{T}_x^i)$ and $\phi(\mathcal{T}_y^j)$ (for $i$ and $j = 1, 2, 3$) thus gives nine cases. With $\alpha_1 := \min_{y\neq x} V(y)$ and $\alpha_2 := \max_{y\neq x} V(y)$, we are sure that when $V(x) \leq 2(3\alpha_1 - 2\alpha_2 - n)$, then $\Phi(x) < \Phi(y)$, for all $y \neq x$.

The construction for this example can be generalized to more general graphs, but we skip the general formulation.

The point of the two examples so far is that we can both with activities and with a potential greatly control what states are most probable in a strongly driven system, as they clearly modify the lifetime of a state. Of course, the dominant states can also be modified by adding edges (accessibilities), as the following example shows.

**Example 3.3** Consider a random walk on a ring $\mathbb{Z}_n$ plus an extra edge, i.e., the ring has a "hair" $\{v, y\}$ where $v \in \mathbb{Z}_n$ and $y$ is outside the ring. The transition rates are taken

$$k(x, x+1) = a_x e^{\mathcal{E}b} e^{\mathcal{E}c}, \quad k(x, x-1) = a_{x-1} e^{\mathcal{E}b} e^{-\mathcal{E}c}, \quad \text{driving } c > 0 \text{ on the ring}$$
$$k(v, y) = \gamma e^{\mathcal{E}\alpha} e^{\mathcal{E}\beta} \quad k(y, v) = \gamma e^{\mathcal{E}\alpha} e^{-\mathcal{E}\beta}, \quad \beta > 0 \text{ on the hair}$$

The results of the computations are summarized in the Table below where $x \neq v$ stands for a generic state on the ring.

| .. | $\Gamma$ | $-\Theta$ | $\Gamma - \Theta$ |
|---|---|---|---|
| $x$ | $-(b+c)$ | $0$ | $-(b+c)$ |
| $v$ | $-\max\{b+c, \alpha+\beta\}$ | $0$ | $-\max\{b+c, \alpha+\beta\}$ |
| $y$ | $-(\alpha-\beta)$ | $-\max\{b+c, \alpha+\beta\} + (\alpha+\beta)$ | $-\max\{b+c, \alpha+\beta\} + 2\beta$ |

The stationary distribution for large $\mathcal{E}$ is up to corrections (for some $\delta > 0$),

$$\rho^s(x) \propto \frac{1}{a_x} e^{-\mathcal{E}(b+c)} (1 + O(e^{-\delta\mathcal{E}})) \text{ on the ring except for}$$

$$\rho^s(v) \propto \frac{1}{a_v} e^{-\mathcal{E}\max\{(b+c),(\alpha+\beta)\}}(1 + O(e^{-\delta\mathcal{E}}))$$

and

$$\rho^s(y) \propto \frac{1}{a_v} e^{-\mathcal{E}(\max\{(b+c),(\alpha+\beta)\}-(\alpha+\beta)+(\alpha-\beta))} (1 + O(e^{-\delta\mathcal{E}}))$$

Note that the choice of parameters $b, c, \alpha, \beta$ will decide the location and the number of dominant states in (3.5).

### 3.2 Response

The linear response for far-from-equilibrium dynamics has some simplifications as well. We focus here on the shift in the stationary distribution when factors are added to the transition rates which are subexponential in $\mathcal{E}$. Naturally, we can use the formula (3.4).

The point of departure is (3.1), where we perturb

$$a(x, y) \to a_\delta(x, y) = a(x, y)(1 + \delta b(x, y))$$

where $\delta \ll 1$ for $b(x, y)$ which are subexponential in $\mathcal{E}$. Quite clearly, $U(x, y)$ and $\Gamma(x)$ do not change; only $A(x) \to A_\delta(x)$ is changing in (3.4):

$$A_\delta(x) = \sum_{\mathcal{T} \in \mathcal{M}(x)} \prod_{(y,y') \in \mathcal{T}_x} a(y, y')(1 + \delta b(y, y'))$$

$$= A(x) + \delta \sum_{\mathcal{T} \in \mathcal{M}(x)} \left( \prod_{(y,y') \in \mathcal{T}_x} a(y, y') \right) \sum_{(y,y') \in \mathcal{T}_x} b(y, y') \quad (3.9)$$

to linear order. Obviously, there is no way to change the dominant states here, but within the dominant states the most probable state may shift by the inclusion of the $b(y, y')$-perturbation.

**Example 3.4** Consider a random walk on $\mathbb{Z}_4$, the $\{1, 2, 3, 4\}$-ring, with transition rates

$$k(x, x \pm 1) = \psi(x, y) e^{\pm \mathcal{E}/2 + \mathcal{E}(v(x) - v(x \pm 1))}$$

where $v(1) = 2$, $v(2) = 3$, $v(3) = 2$, $v(4) = 1$. Using (3.4),

$$A(1) = \psi(2,3)\psi(3,4)\psi(4,1); \quad A(2) = \psi(3,4)\psi(4,1)\psi(1,2)$$
$$A(3) = \psi(4,1)\psi(1,2)\psi(2,3); \quad A(4) = \psi(1,2)\psi(2,3)\psi(3,4) \quad (3.10)$$

and, for some $\tilde{\delta} > 0$,

$$\rho^s(1) \propto A(1)(1 + O(e^{-\tilde{\delta}\mathcal{E}})), \quad \rho^s(2) \propto A(2) e^{-2\mathcal{E}}(1 + O(e^{-\tilde{\delta}\mathcal{E}})),$$
$$\rho^s(3) \propto A(3) e^{-2\mathcal{E}}(1 + O(e^{-\tilde{\delta}\mathcal{E}})), \quad \rho^s(4) \propto A(4)(1 + O(e^{-\tilde{\delta}\mathcal{E}})).$$

With the choice $\psi(1, 2) = \psi(2, 3) = \psi(3, 4) = \psi(4, 1) = 1$, we have that "1" and "4" are most probable (and both are dominant).

Perturbing only over the $\{1, 4\}$-edge, by taking $\psi_\delta(1, 4) = 1 + \delta \alpha$, $\delta \ll 1$, we get $A_\delta(1) = A_\delta(2) = A_\delta(3) = 1 + \delta \alpha$ and $A_\delta(4) = 1$ for (3.10). Then, for $\mathcal{E} \gg 1$, to linear order in $\delta$, the perturbed stationary distribution has (only) "1" as most probable state (between the two dominant states) for $\alpha > 0$ (and "4" as most probable when $\alpha < 0$.). That is of course similar to the selection mechanism in Example 3.1.

The shift in the stationary distribution at large driving has even more important consequences for the currents. In many physical situations, the infinite-driving limit shows a saturation of the current. At that moment, changes in the reactivities, especially when depending on the driving, may have major effects. We know from previous work how phenomena of negative differential susceptibilities or death&resurrection of currents may follow; see e.g. [22, 24, 25]. Here we expand on these topics from the graphical representation (3.4).

We come back to the graphical representation of currents in Sect. 4. Here, we simply use the previous far-from-equilibrium expressions. The (net) stationary current over an edge $x \to y$ is

$$J(x, y) = \rho^s(x) k(x, y) - \rho^s(y) k(y, x)$$
$$= \frac{1}{\mathcal{Z}} \left[ A(x) e^{-\mathcal{E}\Theta(x)} a(x, y) e^{-\mathcal{E}U(x,y)} - A(y) a(y, x) e^{-\mathcal{E}U(y,x)} e^{-\mathcal{E}\Theta(y)} \right] \quad (3.11)$$

where we substituted already the far-from-equilibrium expressions (3.4) and (3.1). The normalization $\mathcal{Z}$ may depend on $\mathcal{E}$ subexponentially, changing the rate of time (scale for the current) but is the same for all edges. Let us assume that for a given choice of rates and for their specific dependence on $\mathcal{E}$, the current saturates to a finite nonzero value. In particular then, assuming that this current is no longer varying exponentially for $\mathcal{E} \uparrow \infty$, we are left with

$$J(x, y) = \frac{1}{\mathcal{Z}} \left[ A(x) a(x, y) - A(y) a(y, x) \right] \quad (3.12)$$

Since the $A$ may still depend on the driving $\mathcal{E}$ (but subexponentially), the saturation of the current is unstable. In particular, changes in the time-symmetric reactivities $\psi_\mathcal{E}(x, y) = \psi_\mathcal{E}(y, x)$ can change the fate of the current at large driving; see again (2.1) and the discussion around (3.1). The magnitude of these changes can be expected to be much larger than is possible close-to-equilibrium, from the dependence on the (large) $\mathcal{E}$. We illustrate with an example.

**Example 3.5** Consider the graph in Fig. 8.

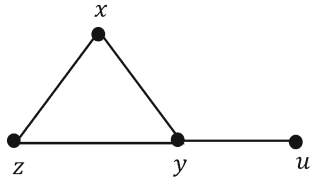

**Fig. 8** Three states $x, y, z$ and a hair $\{y, u\}$

The transition rates are parameterized with $a, b, c, d > 0$ for the symmetric reactivities and driving $\mathcal{E}$:

$$k(x, y) = ae^{2\mathcal{E}}, k(y, z) = be^{2\mathcal{E}}, k(z, x) = c, k(y, u) = de^{\mathcal{E}}$$
$$k(y, x) = ae^{-2\mathcal{E}}, k(z, y) = be^{-2\mathcal{E}}, k(x, z) = c, k(u, y) = de^{-\mathcal{E}}$$

The dominant states are $u$ and $z$.

When the $a, b, c, d$ are kept constant, the current over the edge $(x, y)$ saturates as

$$\lim_{\mathcal{E} \to \infty} J(x, y) = \frac{bc}{b+c}. \tag{3.13}$$

When $b = c = 1$, the current reaches $1/2$ at large driving, which makes the horizontal (red) line in Fig. 9.

However, if we take $a = b = c = d = (\mathcal{E})^{1/3}$ and inspect the current for $\mathcal{E} > 1$, we get the increasing (blue) line in Fig. 9, while to kill the current we can take $a = b = c = d = (\mathcal{E})^{-1/3}$; see the decreasing (green) line in Fig. 9.

Let us now take

$$a_\delta = a(1+\delta), \quad b_\delta = b(1+\delta)$$
$$c_\delta = c(1+\delta), \quad d_\delta = d(1+\delta)$$

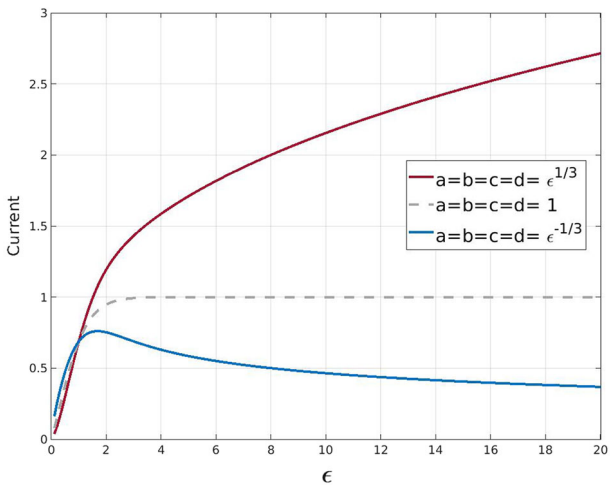

**Fig. 9** The current depends on how the time-symmetric part in the transition rates depends on the driving. That plays little role at small driving, but is governing the current at large driving

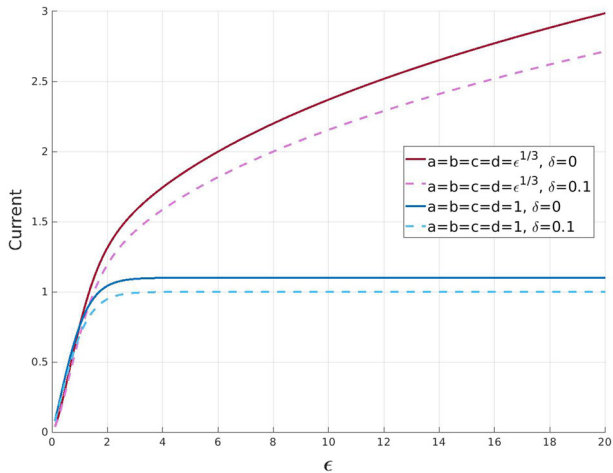

**Fig. 10** The effect of perturbing the symmetric parts on the current

The effect is plotted in Fig. 10.

The increasing (pink and green) lines are diverging from each other, which shows the big effect of $\delta = 0.1$. Of course, when the original $a, b, c, d$ did not depend on $\mathcal{E}$, we see a constant shift in the current as visible in the horizontal (blue and red) lines.

## 4 Currents and Traffic

When breaking detailed balance, stationary currents appear in the graph, as we already had in Example 3.5. They refer to transport and it is interesting to get also for them a graphical representation. A typical application is in deriving the low-temperature behavior. In general the (net) stationary current over the bond $x \rightarrow y$ is

$$J(x, y) = \rho^s(x)k(x, y) - \rho^s(y)k(y, x)$$
$$= \frac{1}{W}\big(w(x)k(x, y) - w(y)k(y, x)\big) \quad (4.1)$$

where $W$ and $w(x)$ are defined in (2.3).

Current is however a dimension-full quantity; it is a flux. A natural time scale is provided by the so called *traffic*, i.e., the time-symmetric current

$$T(x, y) = \rho^s(x)k(x, y) + \rho^s(y)k(y, x) \quad (4.2)$$

Hence and again, for both $J(x, y)$ and $T(x, y)$ the Kirchhoff formula (2.3) for $\rho^s$ can be used in the stationary condition.

### 4.1 One Loop Formula

Consider first graphs as in Fig. 11.

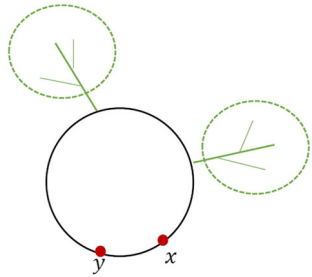

**Fig. 11** Graph with one loop with rooted trees (hairs)

Consider a graph for which the edge $(x, y)$ belongs to a single loop $\ell = \ell_{xy}$. Some vertices on $\ell$ are connected to trees (the *hairs*). In that case, as shown in Appendix B, (4.1) becomes

$$J(x, y) = \frac{W(H_\ell)}{W}(w(\ell^+) - w(\ell^-)) \quad (4.3)$$

where $H_\ell$ is the set of hairs $h$ on the loop and $W(H_\ell)$ is the weight of all hairs $h$ which are towards the loop $\ell$:

$$W(H_\ell) = \prod_{h \in H} \prod_{(z,z') \in h} k(z, z')$$

If a loop does not have hair, then $W(H_\ell) = 1$. Finally, $w(\ell^+)$ is the weight of the loop in the same direction of $(x, y)$ and $w(\ell^-)$ is in the opposite direction. To illustrate some points of formula (4.3), we add the simplest examples.

*Example 4.1* Consider first the graph in Fig. 12.

The spanning trees are

$$\mathcal{T}^1 = \{\{x, y\}, \{x, z\}\}, \quad \mathcal{T}^2 = \{\{x, z\}, \{z, y\}\}, \quad \mathcal{T}^3 = \{\{x, y\}, \{z, y\}\}$$

The normalization in (4.3) is $W = w(x) + w(y) + w(z)$ with

$$\begin{aligned} w(x) &= k_{yx}k_{zy} + k_{zx}k_{yx} + k_{yz}k_{zx} := A \\ w(y) &= k_{zx}k_{xy} + k_{zy}k_{xy} + k_{xz}k_{zy} := B \\ w(z) &= k_{yx}k_{xz} + k_{xy}k_{yz} + k_{xz}k_{yz} := C \end{aligned} \quad (4.4)$$

Furthermore, $w(\ell^+) = k_{xy}k_{yz}k_{zx}$ and $w(\ell^-) = k_{xz}k_{zy}k_{yx}$. As there are no hairs, $W(H_\ell) = 1$. Hence, (4.3) yields

$$J(x, y) = \frac{w(\ell^+) - w(\ell^-)}{A + B + C} \quad (4.5)$$

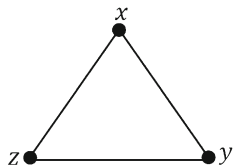

**Fig. 12** A ring with three states $x, y, z$

In order to be more explicit, take $a > 0$ and

$$k(x, y) = a\, e^{\mathcal{E}/2}, k(y, z) = k(z, x) = e^{\mathcal{E}/2}, \quad k(y, x) = a\, e^{-\mathcal{E}/2}, k(z, y) = k(x, z) = e^{-\mathcal{E}/2}$$

The (edge-independent) stationary current over the loop is

$$J_\ell = J(x, y) = \frac{2a \sinh\left(\frac{3}{2}\mathcal{E}\right)}{(4a + 2) \cosh(\mathcal{E}) + a + 1}$$

It increases with $a$ for all $\mathcal{E} \neq 0$. On the other hand, the stationary traffic in the edges is

$$T(x, y) \propto 2a \cosh\left(\frac{3}{2}\mathcal{E}\right) + 4a^2 \cosh\left(\frac{1}{2}\mathcal{E}\right)$$

$$T(y, z) \propto 2a \cosh\left(\frac{3}{2}\mathcal{E}\right) + 2(e^{-\mathcal{E}/2} + ae^{\mathcal{E}/2})$$

$$T(z, x) \propto 2a \cosh\left(\frac{3}{2}\mathcal{E}\right) + 2(ae^{-\mathcal{E}/2} + e^{\mathcal{E}/2})$$

Summing the dynamical activity over the edges in the loop gives the dimensionless ratio

$$\frac{J_\ell}{T_\ell} = \frac{2a \sinh\left(\frac{3}{2}\mathcal{E}\right)}{6a \cosh\left(\frac{3}{2}\mathcal{E}\right) + 4 \cosh\left(\frac{\mathcal{E}}{2}\right)(a^2 + a + 1)}$$

In the $\mathcal{E}$-limit, for $a = e^{\alpha\mathcal{E}}$, the ratio exponentially vanishes for $|\alpha| > 1$, and goes to a constant for $|\alpha| < 1$. That is also the typical scenario, where that ratio is a measure of how many jumps are in the same direction as the current.

**Example 4.2** To illustrate the influence of hair, we next consider the same ring as in the previous example but we add a hair on the state $y$ as in Fig. 8. Using the notation of (4.4),

$$w(x) = Ak_{uy}; \quad w(y) = Bk_{uy} \quad w(z) = Ck_{uy}; \quad w(u) = Bk_{yu}$$

From (4.3), the current over the edge $(x, y)$ becomes

$$J(x, y) = \frac{k_{uy}\left(w(\ell^+) - w(\ell^-)\right)}{k_{uy}(A + B + C) + k_{yu}B} \tag{4.6}$$

We see that putting $k_{yu} = 0$ yields the current in (4.5). In general, to obtain the current (4.5), the rate $k_{yu}$ in (4.6) should go to zero faster than $k_{uy}$. On the other hand, when $k_{uy} = 0$ and $k_{yu} > 0$, then the current in (4.6) is zero. Then in fact, $w(x) = w(y) = w(z) = 0$ implying that all weight goes to state $u$.

### 4.2 General Formula

To move beyond (4.1), we go to the case of two coupled loops for illustrating the scenario.

**Example 4.3** Consider again the graph in Fig. 3 where $(x, y)$ belongs to two loops. Using the tree weights derived in Example 2.2, the current over $(x, y)$ equals

$$J(x, y) = \frac{1}{W} [k(x, y)w(x) - k(y, x)w(y)]$$

$$= \frac{1}{W} \Big[ k_{xy} \big( k_{yz}k_{zu}k_{ux} + k_{uz}k_{zy}k_{yx} + k_{zu}k_{ux}k_{yx} + k_{zy}k_{yx}k_{ux}$$
$$+ k_{yz}k_{zx}k_{ux} + k_{yx}k_{uz}k_{zx} + k_{ux}k_{zx}k_{yx} + k_{yz}k_{zx}k_{uz} \big)$$
$$- k_{yx} \big( k_{xu}k_{uz}k_{zy} + k_{uz}k_{zy}k_{xy} + k_{zu}k_{ux}k_{xy} + k_{ux}k_{zy}k_{xy}$$
$$+ k_{ux}k_{xz}k_{zy} + k_{uz}k_{zx}k_{xy} + k_{ux}k_{zx}k_{xy} + k_{xz}k_{uz}k_{zy} \big) \Big]$$

$$= \frac{1}{W} \Big[ \big( k_{xy}k_{yz}k_{zu}k_{ux} - k_{yx}k_{xu}k_{uz}k_{zy} \big) - (k_{uz} + k_{ux}) \big( k_{xy}k_{yz}k_{zx} + k_{xz}k_{zy}k_{yx} \big) \Big],$$

which can be written as the sum of (4.3) over the two loops,

$$J(x, y) = \frac{1}{W} \Big[ (w(\ell_{\text{larg}}^{x \to y}) - w(\ell_{\text{larg}}^{y \to x})) + (k_{uz} + k_{ux})(w(\ell_{\text{up}}^{x \to y}) - w(\ell_{\text{up}}^{y \to x})) \Big]$$

The edge $\{x, y\}$ belongs to two loops: the largest loop denoted by $\ell_{\text{large}}$ is the loop $x - y - z - u$ and the upper loop $\ell_{\text{up}}$ is the loop $x - y - z$. In general the product of $w(x)k(x, y)$ equals the weight of a new graph created by adding the edge $(x, y)$ to the spanning trees directed toward $x$. Look at Fig. 4: depending on having the edge $\{x, y\}$ or not, the spanning trees come in two groups. There are three spanning trees (the first in the top row, and the first and the last in the bottom row) that do not have the edge $\{x, y\}$. If we add the edge $\{x, y\}$ to these spanning trees, then loops are created. The created loops are exactly $\ell_{\text{large}}$ and $\ell_{\text{up}}$ in the graph Fig. 3. (That idea is used in Appendix B.)

In Appendix B we derive the graphical representation of the current over an edge in the general setting (related to the subject pioneered in [26]), with result

$$J(x, y) = \frac{1}{W} \sum_{\ell_{xy}} \sum_{H_{\ell_{xy}}} W(H_{\ell_{xy}}) \left( w(\ell_{xy}^{x \to y}) - w(\ell_{xy}^{y \to x}) \right) \quad (4.7)$$

Here, $\ell_{xy}$ denotes any loop including $(x, y)$. We sum over all such loops. $H_{\ell_{xy}}$ is the collection of "hairs" towards the loop $\ell_{xy}$. Moreover, $W(H) = 1$ when there are no hairs. In Example 4.3, the largest loop does not have hair and the upper loop has two possible hairs, $k_{uz}$ and $k_{ux}$. We add some examples.

**Example 4.4** Consider Fig. 13, and let $\ell_r$ be the right loop.
The current over the edge $(x, y)$ is

$$J(x, y) = \frac{\sum_{H_{\ell_r}} W(H_{\ell_r})}{W} (w(\ell_r^{x \to y}) - w(\ell_r^{y \to x})) \quad (4.8)$$

where the sum is over all possible hairs created by the sub-graph $G \setminus \ell_r$; see Fig. 14.

Applying the general formula (4.8) gives

$$J(x, y) = \frac{k_{y'z'}k_{z'x'}k_{x'x} + k_{z'y'}k_{y'x'}k_{x'x} + k_{y'x'}k_{z'x'}k_{x'x}}{W} (w(\ell_r^{x \to y}) - w(\ell_r^{y \to x}))$$

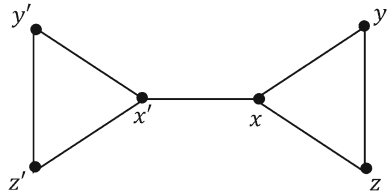

**Fig. 13** Two connected loops

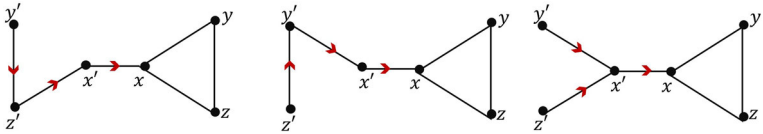

**Fig. 14** Three different hair-structures towards the right loop in Fig. 13

where $w(\ell_r^{x \to y}) = k_{xy}k_{yz}k_{zx}$, and $W$ is

$$W = \Big[(k_{yz}k_{zx} + k_{yx}k_{zx} + k_{zy}k_{yx}) + (k_{xz}k_{zy} + k_{zy}k_{xy} + k_{zx}k_{xy})$$
$$+ (k_{yz}k_{xz} + k_{yx}k_{xz} + k_{xy}k_{yz})\Big](k_{y'z'}k_{z'x'} + k_{y'x'}k_{z'x'} + k_{z'y'}k_{y'x'})k_{x'x}$$
$$+ \Big[(k_{y'z'}k_{z'x'} + k_{y'x'}k_{z'x'} + k_{z'y'}k_{y'x'}) + (k_{x'z'}k_{z'y'} + k_{z'y'}k_{x'y'} + k_{z'x'}k_{x'y'})$$
$$+ (k_{y'z'}k_{x'z'} + k_{y'x'}k_{x'z'} + k_{x'y'}k_{y'z'})\Big](k_{yz}k_{zx} + k_{yx}k_{zx} + k_{zy}k_{yx})k_{xx'}.$$

If we put $k_{xx'} = 0$, while keeping $k_{x'x} \neq 0$, then

$J(x, y)$
$$= \frac{w(\ell_r^{x \to y}) - w(\ell_r^{y \to x})}{(k_{yz}k_{zx} + k_{yx}k_{zx} + k_{zy}k_{yx}) + (k_{xz}k_{zy} + k_{zy}k_{xy} + k_{zx}k_{xy}) + (k_{yz}k_{xz} + k_{yx}k_{xz} + k_{xy}k_{yz})}$$

which equals the current over a single triangle with states $\{x, y, z\}$. In all cases, the current over the edge $\{x, x'\}$ is zero, see (4.7), as the edge is not located on any loop.

**Example 4.5** Consider Fig. 15 where the edge $(x, y)$ is now located on two loops.

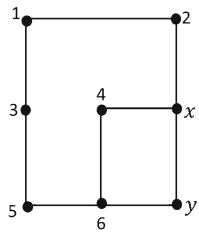

**Fig. 15** Two loops with common edges

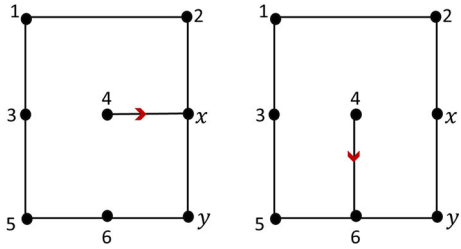

**Fig. 16** The large loop and its hairs

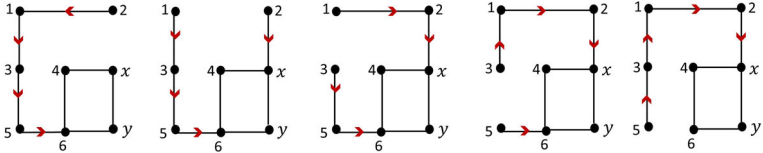

**Fig. 17** The small loop and its hairs

The strategy here is to first find the current in the large loop without considering the small loop; see Fig. 16

$$J(x, y)_{\text{larg}} = \frac{k(4, x)k(4, 6)}{W} (w(\ell_{\text{larg}}^{x \to y}) - w(\ell_{\text{larg}}^{y \to x}))$$

where $w(\ell_{\text{larg}}^{x \to y}) = k_{12}k_{2x}k_{xy}k_{y6}k_{65}k_{53}k_{31}$. Next, consider the small loop and all different possibilities of hairs, as in Fig. 17.

There, the current over $(x, y)$ is

$$J(x, y)_{\text{small}} = \frac{\sum_{H_{\text{small}}} W(H_{\text{small}})}{W} (w(\ell_{\text{small}}^{x \to y}) - w(\ell_{\text{small}}^{y \to x}))$$

where

$$\sum_{H_{\text{small}}} W(H_{\text{small}}) = k_{21}k_{13}k_{35}k_{56} + k_{2x}k_{13}k_{35}k_{56}$$

$$+ k_{2x}k_{12}k_{35}k_{56} + k_{2x}k_{12}k_{31}k_{56} + k_{2x}k_{12}k_{31}k_{53}$$

Finally, the (true) current is

$$J(x, y) = J(x, y)_{\text{large}} + J(x, y)_{\text{small}}$$

illustrating again (4.7).

For the traffic, the analogue of (4.7) is

$$T(x, y) = \frac{1}{W} \sum_{\ell_{xy}} \sum_{H_{\ell_{xy}}} W(H_{\ell_{xy}}) \left( w(\ell_{xy}^{x \to y}) + w(\ell_{xy}^{y \to x}) \right) \qquad (4.9)$$

That closely resembles 4.7: $\ell_{xy}$ denotes a loop including $(x, y)$, and $H_{\ell_{xy}}$ is the collection of hairs on the loop $\ell_{xy}$. Notice that two different directions of an edge are making a loop. For example for the edge $(x, y)$, there is the loop $x - y - x$. In the current formula, the weights of such (trival) loops (made by the two directions of an edge) vanish together, but they do contribute for the traffic. It implies that the sum over all spanning trees towards the edge $(x, y)$ is still there: in contrast with the formula for the current, traffic depends on the weight of the edges coming to $\{x, y\}$.

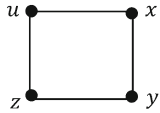

**Fig. 18** A ring with four states

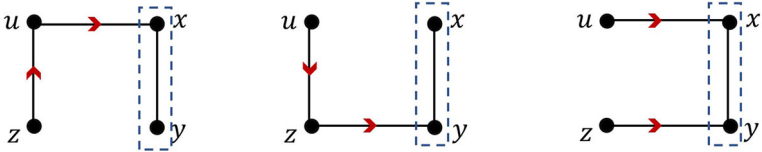

**Fig. 19** Spanning trees towards the edge $x - y$

**Example 4.6** To illustrate the general formula (4.9), we consider a ring with four states $\{x, y, z, u\}$; see Fig. 18.

The spanning trees towards the edge $x - y$ are shown in Fig. 19.

We use (4.9) and the traffic in the edge $x - y$ is

$$T(x, y) = \frac{1}{W} \left( k_{xy} k_{yz} k_{zu} k_{ux} + k_{yx} k_{xu} k_{uz} k_{zy} \right) + \frac{k_{zu} k_{ux} + k_{uz} k_{zy} + k_{ux} k_{zy}}{W} \left( 2 k_{xy} k_{yx} \right)$$

where we inserted the weight of the three spanning trees towards $x - y$.

## 5 Close to Equilibrium: Tree–McLennan Formula

The relation (2.9) is valid without further assumptions. We can derive from it a tree-McLennan relation for the close-to-equilibrium regime, by considering transition rates

$$k(x, y) = k_{\text{eq}}(x, y) \, e^{\frac{\varepsilon}{2} s(x, y)} \tag{5.1}$$

Here $k_{\text{eq}}(x, y)$ is the detailed balance (equilibrium) part,

$$\frac{k_{\text{eq}}(x, y)}{k_{\text{eq}}(y, x)} = e^{V(x) - V(y)}$$

(for some potential $V$ and putting the inverse temperature $\beta = 1$) and the $s(x, y)$ are antisymmetric with a small amplitude $\varepsilon$. A McLennan formula expresses the stationary close-to-equilibrium distribution $\rho^s = \rho^s_\varepsilon$ in terms of the associated equilibrium one, $\rho_{\text{eq}}(x) = \rho^s_0(x) \propto e^{-V(x)}$. That McLennan formula can be written as

$$L_{\text{eq}} \left( \frac{\rho^s(x)}{\rho_{\text{eq}}(x)} - 1 \right) = \varepsilon \sum_z k_{\text{eq}}(x, z) s(x, z) + O(\varepsilon^2) \tag{5.2}$$

where $L_{\text{eq}} g(x) = \sum_z k_{\text{eq}}(x, z)(g(z) - g(x))$ is the backward generator of the equilibrium process on a function $g$; see [27–29]. Or, to first order $\varepsilon$ in the driving, the expected dissipated power under the equilibrium dynamics is a pseudo-potential for the close-to-equilibrium stationary distribution; see more in Sect. 6.

To see its arborification, we note first that the writing (5.1) naturally gives rise to

$$S_{\mathcal{T}}(y \to x) = V(y) - V(x) + \varepsilon S^1_{\mathcal{T}}(y \to x)$$

where $S^1_{\mathcal{T}}(y \to x)$ is the sum of the entropy fluxes $s(u, v)$. Hence, from (2.9),

$$\frac{\rho^s_\varepsilon(x)}{\rho^s_0(x)} = 1 + \varepsilon \left\langle \langle S^1_{\mathcal{T}}(y \to x) \rangle \right\rangle^{\text{eq}}_x + O(\varepsilon^2) \quad (5.3)$$

The double expectation $\langle\langle \cdot \rangle\rangle^{\text{eq}}_x$ is over the equilibrium distribution (sum with $\rho^s_0(y)$) and over the equilibrium tree-ensemble (2.7). Equation (5.3) is a tree-McLennan formula for the close-to-equilibrium ensemble.

We can be more explicit here and show that (5.3) exactly reproduces the (standard) McLennan formula (5.2). We skip the many details but the main point is that

$$\sum_z k_{\text{eq}}(x, z) \sum_{\mathcal{T}} W_{\text{eq}}(\mathcal{T}_x) S^1_{\mathcal{T}}(x \to z)) = w_{\text{eq}}(x) \sum_z k_{\text{eq}}(x, z) s^1(x, z)$$

where the tree-representation in the left-hand side gets connected with the dissipated power in the right-hand side. More technical details are given in Appendix A.

## 6 Quasi-potential

Nonequilibrium conditions are not governed by balancing energy and entropy as it happens in equilibrium statistical mechanics. Similarly, excess heat in the quasistatic relaxation between different equilibrium conditions is not given in terms of a change in energy of the system. In nonequilibrium that gets replace by a quasi-potential that measures the additional heat that flows to a thermal bath; see e.g. [30–33].

The definition has already been encountered before. Looking back at the McLennan formula (5.2), we see an equation of the form

$$LV = -f \quad (6.1)$$

where $L = L_{\text{eq}}$ is the equilibrium generator, $V(x) = \rho^s(x)/\rho_{\text{eq}}(x) - 1$ and $f(x) = -\varepsilon \sum_z k_{\text{eq}}(x, z) s(x, z)$, where $\langle V \rangle_{\text{eq}} = 0$ (because $\sum_x \rho_{\text{eq}}(x)[\rho^s(x)/\rho_{\text{eq}}(x) - 1] = 0$) and $\langle f \rangle_{\text{eq}} = 0$ (because $\sum_{x,z} \rho_{\text{eq}}(x) k_{\text{eq}}(x, z) s(x, z) = 0$). Such an equation, where $f$ is given with expectation $\langle f \rangle = 0$ in the stationary distribution as defined from the jump process with generator $L$, asks for finding the so called quasi-potential $V$, again with $\langle V \rangle = 0$. The solution is given by the excess

$$V_f(x) = \int_0^\infty dt \, e^{tL} f(x) = \int_0^\infty \langle f(X_t) \mid X_0 = x \rangle \, dt \quad (6.2)$$

Obviously, this equation can as well serve as definition of quasi-potential, and is not restricted to the close-to-equilibrium regime. It is an excess quantity (transient versus stationary) and the good news is that there is a graphical representation for $V$, and hence for integrals like (6.2) which are important for computing excess work or excess heat, [34, 37]. The representation (5.3) and the identities derived in Appendix A are exactly a manifestation of that, but are restricted to the close-to-equilibrium regime. Here we highlight the more general case for solving (6.1) and that is done using the so called matrix-forest theorem, an extension indeed of the matrix-tree theorem which implies the Kirchhoff formula (2.3).

### 6.1 Matrix-Forest Theorem

It turns out (a) that a proper meaning can be given to inverting (6.1) via the resolvent, taking

$$V_f = \lim_{\alpha \uparrow \infty} \alpha \frac{1}{1 - \alpha L} f \qquad (6.3)$$

and (b) that there is a graphical representation for that resolvent expression, which falls under the so called matrix-forest theorem, [38, 39]. Here we present the upshot of those results as concerns the computation of (6.3) and hence of (6.2). Applications are found in [16, 35–37], where the source $f$ is the excess power, which leads to the interpretation of $V_f$ as the excess heat along the relaxation to a (new) steady state from different initial configurations. Similarly, we can define excess work.

We introduce the main aspects of formulation.
The right-hand side of (6.3) can be written using the matrix-forest theorem. A forest is a set of (disjoint) trees, including single vertices. When the graph has $n$ vertices, then

$$V_f(x) = \sum_y \frac{w(\mathcal{F}_{n-2}^{x \to y})}{W} f(y) \qquad (6.4)$$

The notation uses $\mathcal{F}_{n-2}^{x \to y}$ for the set of all forests with $n-2$ edges (i.e., consisting of just two trees), where there is no edge going out from $y$ and where there is a path from $x$ to $y$. A special case is $\mathcal{F}_{n-2}^{x \to x}$ is the set of all forests with $n-2$ edges with no edge going out from $x$. The weight is

$$w(\mathcal{F}) = \sum_{F \in \mathcal{F}} \prod_{(z,z') \in F} k(z, z'),$$

product over all forests. Here comes an example.

**Example 6.1** Consider the graph $\mathbb{Z}_3$. The set $\mathcal{F}_1^{x \to y} = \{\{z\}, \{(x, y)\}\}$ has just one forest containing two trees.

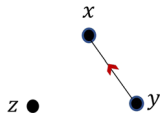

On the other hand, $\mathcal{F}_1^{x \to x}$ contains

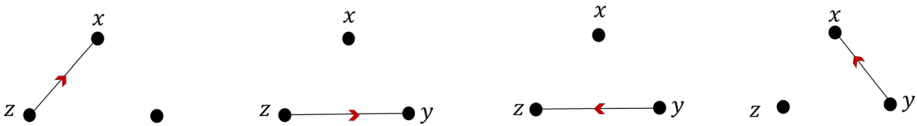

The quasi-potential is

$$V_f(x) = \frac{w(\mathcal{F}_1^{x \to x}) f(x) + w(\mathcal{F}_1^{x \to y}) f(y) + w(\mathcal{F}_1^{x \to z}) f(z)}{W}$$

where

$$w(\mathcal{F}_1^{x \to x}) = k(z, x) + k(z, y) + k(y, z) + k(y, x)$$
$$w(\mathcal{F}_1^{x \to y}) = k(x, y)$$
$$w(\mathcal{F}_1^{x \to z}) = k(x, z).$$

At large driving we just have to use the digraph of preferred successors; see (3.2) and the discussion following it. Assuming that $\Gamma(i) + U(i, i+1) < 0$, $\Gamma(i) + U(i, i-1) > 0$, we get

$$V_f(x) \propto [k(z, x) + k(y, z)] f(x) + k(x, y) f(y)$$

as $\mathcal{E} \gg 1$.

### 6.2 Large Driving

Just as for the Kirchhoff formula in Sect. 3, we get a simplification in the application of the matrix-forest theorem when the driving $\mathcal{E}$ is large. Our main remark here is that an exchange of limits (large driving and infinite time in (6.3)) is not always allowed. The problem is that the denominator $W$ in (6.4) may go to zero, while some of the forest weights remain bounded. That of course reflects the fact that the relaxation time (and hence the convergence in (6.2)) may not be uniformly bounded in $\mathcal{E}$. The system can get trapped in a region of the graph for a much longer time compared to the life-time of dominant states.

We give an example and a counterexample.

**Example 6.2** For a random walker on $\mathbb{Z}_N$, we consider the rates

$$k_{\mathcal{E}}(x, x \pm 1) = k_0(x, x \pm 1) e^{\pm \mathcal{E}/2}$$

where the $k_0(x, x \pm 1) > 0$ are arbitrary. Fixing any function $f$, we can find the quasi-potential $V_{f,\mathcal{E}}$ from (6.4). The same thing can be done for the rates

$$k_{\infty}(x, x+1) = k(x, x+1), \qquad k_{\infty}(x, x-1) = 0$$

and again the quasi-potential $V_{f,\infty}$ can be constructed from (6.4). Note that upon rescaling time by $e^{\mathcal{E}/2}$ the transition rates become bounded and

$$\lim_{\mathcal{E}} e^{-\mathcal{E}/2} k_{\mathcal{E}}(x, x \pm 1) = k_{\infty}(x, x \pm 1)$$

and the resulting graph with oriented edges remains irreducible.
It is then easy to check that $\lim_{\mathcal{E}} V_{f,\mathcal{E}} = V_{f,\infty}$.

**Example 6.3** Consider again the graph in Fig. 1. We take specific rates

$$k_{\mathcal{E}}(x, y) = k_{\mathcal{E}}(y, z) = k_{\mathcal{E}}(x, z) = k_{\mathcal{E}}(u, y) = k_{\mathcal{E}}(u, w) = 1,$$
$$k_{\mathcal{E}}(z, y) = k_{\mathcal{E}}(y, x) = e^{-\mathcal{E}}, k_{\mathcal{E}}(y, u) = e^{-7\mathcal{E}}, k_{\mathcal{E}}(z, x) = e^{-2\mathcal{E}}, k_{\mathcal{E}}(w, u) = e^{-5\mathcal{E}}$$

Here $W \simeq e^{-6\mathcal{E}}$ goes to zero. Taking the function $f(i) = \delta_{i,w}$ which is nonzero (equal to one) only at the site $w$, we have from (6.4) that

$$V_{f,\mathcal{E}}(w) \geq \frac{w(\mathcal{F}_3^{w \to w})}{W}$$

and in $\mathcal{F}_3^{w \to w}$ there is the forest with the tree $(x \to z, u \to y \to z)$ and the singleton $w$, which has weight 1. Therefore here, $V_{f,\mathcal{E}}(w)$ goes to infinity with $\mathcal{E}$.

Large driving is obviously related to driving under low-temperature conditions. More applications will be presented elsewhere; see e.g. [16, 37].

## 7 Conclusions

Arborification as implementation of matrix-tree and matrix-forest theorems, leads to new graphical ensembles that present alternatives to path-space integration. The result are graphical representations of stationary distributions, current, traffic and quasi-potentials where possibilities appear for purposes of nonequilibrium statistical mechanics. Certain features become especially clear in asymptotic regimes such as at low temperature, at small and at large driving or for special graphs.

## Declarations

**Conflicts of interest** The authors have no competing interests to declare that are relevant to the content of this article.

## Appendix A: More on Deriving the McLennan Formula

The first point is that by expanding $w(x)$ up to linear order in $\varepsilon$,

$$\frac{\rho_\varepsilon^s(x)}{\rho_0^s(x)} = 1 + \frac{\varepsilon}{2}\left(\frac{\sum_\mathcal{T} \omega_0(\mathcal{T}_x) S^1(\mathcal{T}_x)}{w_0(x)} - \frac{\sum_y \sum_\mathcal{T} \omega_0(\mathcal{T}_y) S^1(\mathcal{T}_y)}{\sum_y w_0(y)}\right) + O(\varepsilon^2) \quad (A.1)$$

and using 2.5,

$$\frac{\sum_y \sum_\mathcal{T} \omega_0(\mathcal{T}_y) S^1(\mathcal{T}_y)}{\sum_y w_0(y)} = \frac{\sum_y \sum_\mathcal{T} \frac{k_{eq}(x,y)}{k_{eq}(y,x)} \omega_0(\mathcal{T}_x) S^1(\mathcal{T}_x)}{\sum_y \frac{k_{eq}(x,y)}{k_{eq}(y,x)} w_0(x)} + 2 \frac{\sum_y \sum_\mathcal{T} \omega_0(\mathcal{T}_y) S_\mathcal{T}^1(y \to x)}{W_0}$$

$$= \frac{\sum_\mathcal{T} \omega_0(\mathcal{T}_x) S^1(\mathcal{T}_x)}{w_0(x)} + 2 \sum_y \frac{w_0(y)}{W_0} \sum_\mathcal{T} \frac{\omega_0(\mathcal{T}_y) S_\mathcal{T}^1(y \to x)}{w_0(y)}$$

We now get

$$L_{eq}\left(\frac{\rho_\varepsilon^s}{\rho_0^s} - 1\right) = \varepsilon \left(\sum_z k_{eq}(x,z) \frac{\sum_\mathcal{T} \omega_{eq}(\mathcal{T}_x) S_\mathcal{T}^1(x \to z))}{w_{eq}(x)}\right)$$

for the action of the equilibrium backward generator $L_{eq}$.

## Appendix B: Proof of the Current Formulas

We need two lemmas.

**Lemma B.1** *Let $\mathcal{T}^G$ as a set of all spanning trees in graph $\mathcal{G}$ which is a loop and hairs, and $\mathcal{T}^\ell$ as a set of all spanning trees on the loop $\ell$. Then for every vertex $x$ on the loop*

$$\forall \tau^G \in \mathcal{T}^G; \quad \exists! \tau^\ell \in \mathcal{T}^\ell; \quad w(\tau_x^G) = w(\tau_x^\ell) W(H_\ell)$$

*Where $\tau_x^G$ is a spanning tree on the graph $\mathcal{G}$ towards $x$, $\tau_x^\ell$ is a spanning tree in the loop $\ell$ towards $x$ and $W(H_\ell)$ is the weight of all hairs towards the loop.*

The point of that lemma is that the weights of spanning trees (toward $x$ located on the loop) have a common term $W(H)$ which does not depend on $x$.

**Proof** Let $\tau^G$ be a spanning tree in the graph $\mathcal{G}$. It contains all vertices of $\mathcal{G}$, on hairs and loop. The hairs themselves are trees, so that the set of edges of $\tau^G$ includes all edges on the hairs $E(H)$ and

$$E(\tau^\ell) = E(\tau^G) \setminus E(H)$$

with, for every $x \in \mathcal{V}(\ell)$,

$$w(\tau_x^\ell) = \frac{w(\tau_x^G)}{W(H)}$$

$\square$

**Lemma B.2** *For every $x$ and $y \in \mathcal{V}(\ell)$,*

$$\sum_{\tau^\ell} \left( w(\tau_x^\ell) k(x,y) - w(\tau_y^\ell) k(y,x) \right) = w(\ell^{x \to y}) - w(\ell^{y \to x})$$

**Proof** Let $\mathcal{T}_x^\ell$ be the set of all spanning trees on the loop $\ell$ toward $x$. $\mathcal{T}_x^\ell = A_1 \cup A_2$ can be split into two subsets

$$A_1 := \{\tau_x^\ell; (y,x) \in \tau_x^\ell\}, \quad A_2 := \{\tau_x^\ell; (y,x) \notin \tau_x^\ell\}$$

where $\tau_x^\ell \in \mathcal{T}_x^\ell$. Similarly, $\mathcal{T}_y^\ell$ is the set of all spanning trees toward $y$ on the loop and

$$\mathcal{T}_y^\ell = B_1 \cup B_2; \quad B_1 := \{\tau_y^\ell; (x,y) \in \tau_y^\ell\}; \quad B_2 := \{\tau_y^\ell; (x,y) \notin \tau_y^\ell\}$$

$\tau_x$ is a tree towards $x$ so it does not have the edge $(x,y)$ which is leaving the vertex $x$. $\tau_y$ also does not have the edge $(y,x)$. So then

$$\forall \tau_x^\ell \in A_1 \quad \tau_x^\ell \setminus (y,x) \cup (x,y) \in B_1$$

so that

$$\exists \tau_y^\ell \in B_1; \quad \tau_y^\ell = \tau_x^\ell \setminus (y,x) \cup (x,y)$$

In other words

$$w(\tau_x^\ell \setminus (y,x) \cup (x,y)) = w(\tau_x^\ell) \times \frac{k(x,y)}{k(y,x)} = w(\tau_y^\ell)$$

and

$$\forall \tau_y^\ell \in B_1 \quad \tau_y^\ell \setminus (x,y) \cup (y,x) \in A_1$$

so that

$$\exists \tau_x^\ell \in A_1; \quad \tau_x^\ell = \tau_y^\ell \setminus (x,y) \cup (y,x)$$

which means

$$w(\tau_y^\ell \setminus (x,y) \cup (y,x)) = w(\tau_y^\ell) \times \frac{k(y,x)}{k(x,y)} = w(\tau_x^\ell)$$

Hence

$$\sum_{\tau \in A_1} w(\tau_x) k(x,y) = \sum_{\tau \in B_1} w(\tau_y) k(y,x) \tag{B.1}$$

On the other hand

$$\forall \tau_x^\ell \in A_2 \quad \tau_x^\ell \cup (x, y) = \ell^{x \to y}$$
$$\forall \tau_y^\ell \in B_2 \quad \tau_y^\ell \cup (y, x) = \ell^{y \to x}$$

so then

$$w(A_2)k(x, y) = w(\ell^{x \to y}); \quad w(B_2)k(y, x) = w(\ell^{y \to x}) \tag{B.2}$$

Now for every $x$ and $y$ on the loop

$$\sum_{\tau^\ell} \left( w(\tau_x^\ell)k(x, y) - w(\tau_y^\ell)k(y, x) \right) = \sum_{\tau^\ell} w(\tau_x^\ell)k(x, y) - \sum_{\tau^\ell} w(\tau_y^\ell)k(y, x)$$
$$= \left( \sum_{\tau^\ell \in A_1} w(\tau_x^\ell)k(x, y) - \sum_{\tau^\ell \in B_1} w(\tau_y^\ell)k(y, x) \right)$$
$$+ \left( \sum_{\tau^\ell \in A_2} w(\tau_x^\ell)k(x, y) - \sum_{\tau^\ell \in B_2} w(\tau_y^\ell)k(y, x) \right)$$
$$= w(\ell^{x \to y}) - w(\ell^{y \to x})$$

Where in the second line we used B.1 and in the last line we used B.2. □

Next we prove formula 4.3.

**Proof** Consider on the graph $\mathcal{G}$ for every $x$ and $y$ on the loop

$$j(x, y) = \rho^s(x)k(x, y) - \rho^s(y)k(y, x)$$
$$= \frac{w(x)}{W}k(x, y) - \frac{w(y)}{W}k(y, x)$$

where

$$w(x) = \sum_{\tau^G} \prod_{(z, z') \in \tau_x^G} k(z, z') = \sum_{\tau^G} w(\tau_x^G)$$

From Lemma B.1

$$w(x) = W(H) \sum_{\tau^\ell} w(\tau_x^\ell)$$

In the same way

$$w(y) = W(H) \sum_{\tau^\ell} w(\tau_y^\ell)$$

so then

$$j(x, y) = \frac{W(H)}{W} \sum_{\tau^\ell} \left( w(\tau_x^\ell)k(x, y) - w(\tau_y^\ell)k(y, x) \right)$$

From the Lemma B.2 the relation (4.3) has been proved. □

We need another lemma to prove the general formula (4.7).

**Lemma B.3** *Consider a spanning tree $\mathcal{T}$ in the connected graph $\mathcal{G}$. If the edge $\{x, y\}$ is in the graph but is not in the spanning tree, then by adding the edge $\{x, y\}$ to the spanning tree one loop is created.*

**Proof** There is a path between $x$ and $y$ in the spanning tree $\mathcal{T}$. The edge $\{x, y\}$ is a second path between vertices $x$ and $y$. If we have more than one path between two vertices in a graph then we have a loop.

Imagine a general connected graph $\mathcal{G}$ in which the edge $\{x, y\}$ belongs to more than one loop. Corresponding to the edge $\{x, y\}$, the set of all spanning trees $\mathcal{T}^G$ can be split into two sets. The set denoted by $T^{xy}$ is a set of trees including the edge $\{x, y\}$, while the set denoted by $T^c$ is not including the edge $\{x, y\}$. In (B.1) it is shown that

$$\sum_{\mathcal{T} \in T^{xy}} \big( w(\mathcal{T}_x)k(x, y) - w(\mathcal{T}_y)k(y, x) \big) = 0$$

Notice that in the lemma we considered all spanning trees on a loop but the result of (B.1) is the same for every spanning tree.

We focus on the set which does not include the edge $\{x, y\}$. By adding the edge $\{x, y\}$ to the set $T^c$, a loop including this edge is created (Lemma B.3). For every tree of $T^c$, by adding this edge, a loop with a structure of hairs appears. In other words,

$$w(x)k(x, y) = \sum_{\mathcal{T} \in T^c} w(\mathcal{T}_x)k(x, y) = \sum_{\ell_{xy}} \sum_{H_{\ell_{xy}}} w(\ell_{xy}) w(H_{\ell_{xy}})$$

where $\ell_{xy}$ is a loop in the graph $\mathcal{G}$ including the edge $\{x, y\}$ and $H_{\ell_{xy}}$ is one possible structure of hairs connected to the loop $\ell_{xy}$. From here and the relation (4.3), the general formula (4.7) is proved. □